\documentclass{elsart5p}
\usepackage{graphicx}
\usepackage{amssymb}
\usepackage{dcolumn}
\usepackage{bm}

\begin{document}
\begin{frontmatter}

\title{Infrared Investigation of the Charge Ordering Pattern in the Organic
Spin Ladder Candidate (DTTTF)$_2$Cu(mnt)$_2$}

\author{J.~L. Musfeldt$^1$, S. Brown$^{1,*}$, S. Mazumdar$^2$, R.~T. Clay$^3$,
M. Mas-Torrent$^4$, C. Rovira$^4$, J.~C. Dias$^5$}
\author{R.~T. Henriques$^5$, M. Almedia$^5$}
\address{$^1$Department of Chemistry, University of Tennessee,
Knoxville, TN 37996, USA}
\address{$^2$Department of Physics, University of Arizona, Tuscon,
AZ 85721, USA}
\address{$^3$Dept. of Physics and Astronomy, HPC$^2$ Center
for Computational Sciences, Mississippi State University,
Mississippi State, MS 39762, USA}
\address{$^4$Institut de Ci\'encia de Materials de Barcelona, CSIC
Campus de la UAB, E-08193 Bellaterra, Spain}
\address{$^5$Departamento de Quimica,  Instituto Tecnol\'ogico e
Nuclear/Centero de Fisica da Mat\'eria Condensada da Universidade de
Lisboa, P-2686-953 Sacav\'em, Portugal}

\date{\today}

\begin{abstract}

We measured the variable temperature infrared response of the spin
ladder candidate (DTTTF)$_2$Cu(mnt)$_2$ in order to
%
distinguish between two competing ladder models, rectangular versus
zigzag, proposed for this family of materials.
The  distortion along the
stack direction
below  235 K is consistent with
a doubling along $b$ through the metal-insulator
transition.
While this would agree with either of the ladder models,
the concomitant transverse distortion
rules out the rectangular ladder model and
supports the zigzag scenario.
Intramolecular distortions within the DTTTF building block molecule
also give rise to on-site charge asymmetry.

\end{abstract}

\end{frontmatter}

\begin{figure}[t]
\centerline{
\includegraphics[width = 3.25 in]{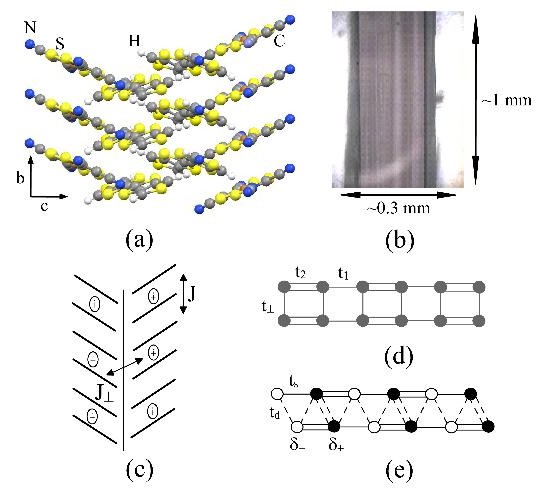}}
\caption{\label{models} (Color online) (a) Crystal structure of
(DTTTF)$_2$Cu(mnt)$_2$  
\cite{Ribas2005}, (b) photograph of the (DTTTF)$_2$Cu(mnt)$_2$
single crystal and the aperture used during our infrared
measurements, (c) proposed low-temperature arrangement of charge
supporting the rectangular spin ladder picture \cite{Ribas2005}, (d)
schematic of the bond distortion pattern within the rectangular
ladder model \cite{Yan2006}, and (e) schematic charge ordering and
bond distortion pattern within the zigzag ladder model.
The grey circles in (d) denote site charge of 0.5 while the filled
(unfilled) circles in (e) denote site charges greater than (less
than) 0.5 \cite{Clay2005}. }
\end{figure}

Quantum spin ladders have attracted considerable interest as
intermediaries between one-dimensional chains and two-dimensional
square lattices \cite{Dagotto1996,Dagotto1999,Rovira2001,Ribas2005}.
Additional interest has arisen from theoretical studies
which find that hole-doped spin ladders can support
superconductivity \cite{Dagotto1999,Maekawa1996}. Whereas most
systems are structural ladders,
organic ladder-like compounds
such as dithiophentelrathiafulvalene copper maleonitrile dithiolate,
(DTTTF)$_2$Cu(mnt)$_2$ are formed by the coupling of molecular
building blocks (Fig. \ref{models}(a)).
This system is
particularly attractive
because it belongs to a family  of quasi-isostructural compounds,
with tunable properties depending on the counterion ($M$ = Pt, Cu,
Au, Ni) \cite{Rovira2001,Ribas2005}. That the cation stacks of
(DTTTF)$_2$Cu(mnt)$_2$ have
a $\frac{1}{4}$-filled band 
with delocalized charge and spin at high temperatures, and are
therefore described by an {\it electronic} extended Hubbard
Hamiltonian, as opposed to the localized Heisenberg spin
Hamiltonian, brings additional complexity to the field.

Two different theoretical models, the rectangular
\cite{Ribas2005,Yan2006,Ribas2005b,Wesolowski2003}  and the zigzag
ladder \cite{Clay2005},  have been proposed for this family of
coupled stack materials (Figs.~\ref{models}(d) and (e))
Although determination of the wavevector dependence of the magnon
bands \cite{Barnes2003} is the strongest test of spin ladder
character and can in principle distinguish between the models,
inelastic neutron scattering experiments on (DTTTF)$_2$Cu(mnt)$_2$
are limited by sample quantity and isotopic decoration requirements
\cite{Sample}. The cations have homogeneous charge within the
rectangular model and are charge ordered within the zigzag model
\cite{Clay2005}. An alternate approach to distinguishing between the
models involves detailed comparison of vibrational property
measurements with theoretical predictions.

(DTTTF)$_2$Cu(mnt)$_2$ is a
donor-acceptor salt that
displays the characteristic herringbone chain structure of many
organic molecular solids (Fig. \ref{models}(a)). 
The $\frac{1}{4}$-filled band donor stack is weakly metallic at room
temperature with $\sigma_{dc}$ $\sim$ 12 $\Omega^{-1}$ cm$^{-1}$
\cite{Dias2004}. (DTTTF)$_2$Cu(mnt)$_2$ has a sharp second-order
metal-insulator transition at 235 K, which doubles the unit cell
along the stacking direction, $b$ \cite{Dias2004,Ribas2005}.
Within the rectangular ladder model,
each dimer unit cell of (DTTTF)$_2$$^{+.}$ radical cations along the
stacking $b$-axis is thought to act as a single site with spin S =
$\frac{1}{2}$ below this transition temperature
\cite{Rovira1997,Rovira2001,Ribas2005}.
The rungs of the proposed ladder lie along $c$ and are formed by
close S$^{...}$S contacts between DTTTF molecules in adjacent
stacks. From susceptibility measurements, magnetic exchange
strengths are estimated to be J$_{\parallel}$ $\sim$ 121 K and
J$_{\perp}$ $\sim$ 218 K \cite{Ribas2005}. The ratio of
J$_{\perp}$/J$_{\parallel}$ is 1.8, indicative of intermediate
coupling. The spin gap is 130 K \cite{Ribas2005}.
Interestingly, application of pressure recovers the metallic state
\cite{Dias2004}. The structural, magnetic, and electrical properties
of doped DTTTF-based materials were also studied, \cite{Ribas2005b}
but superconductivity was not observed.

In order to investigate the charge ordering and bond distortion
patterns (if any) of (DTTTF)$_2$Cu(mnt)$_2$ below the
metal-insulator transition, we  measured the temperature-dependent
infrared vibrational properties of this material. We compare the
results with those of (DTTTF)$_2$Au(mnt)$_2$
\cite{Wesolowski2003} and with theoretical predictions of
intermolecular charge ordering in the rectangular and zigzag ladders
\cite{Ribas2005,Yan2006,Clay2005}.
We find that the zigzag ladder model provides a more appropriate
description of the large transverse structural distortion that
accompanies the 235 K metal-insulator transition.
A mode
analysis provides the
microscopic basis for this distortion. 
We briefly discuss the consequences of intramolecular distortions
for charge ordering within the DTTTF building block molecule in both
the charge transfer salt and neutral molecular solid.


Single crystals of (DTTTF)$_2$Cu(mnt)$_2$  were grown as described
previously \cite{Rovira2001,Ribas2005}. Typical dimensions were 2
$\times$ 0.3 $\times$ 0.1 mm. Our work was done on the largest
($bc$) crystal face. As shown in Fig. \ref{models}(b), the crystal
surfaces were somewhat striated. Variable temperature infrared
reflectance measurements were carried out on these small crystals
using a Bruker 55 Fourier transform infrared spectrometer and
attached microscope accessory. Both wire grid and polaroid film
polarizers were used, as appropriate. Measurements were taken at
several different temperatures using a cryostat and temperature
controller setup, concentrating on the temperature range near the
phase transitions. The cooling rate was slow ($\sim$ a few K/min) to
avoid breakage. Careful aperaturing minimized the impact of surface
quality and small crystal size (Fig. \ref{models}(b)). A
Kramers-Kronig analysis was tried in an attempt to relate the
measured reflectance to the optical constants of the material. In
many cases, the procedure worked well, but in a few cases, less
reasonable results were obtained. We therefore elect to present and
discuss the absolute reflectance spectra of (DTTTF)$_2$Cu(mnt)$_2$,
which captures the important physics of this system over the full
temperature range of investigation without the variability of a
Kramers-Kronig analysis.


The 300 K optical properties of (DTTTF)$_2$Cu(mnt)$_2$  are highly
anisotropic and  similar to those of the Au analog
\cite{Wesolowski2003}, with a combination of electron-molecular
vibrational  and infrared active modes of $B_u$ symmetry plus  a
strong charge transfer band in the chain direction. Along $b$, the
largest modes appear at $\sim$770 and  1300 cm$^{-1}$ in the
reflectance spectrum. We attribute these features to terminal
\begin{figure}[tb]
\centerline{
  \includegraphics[width = 3.6 in]{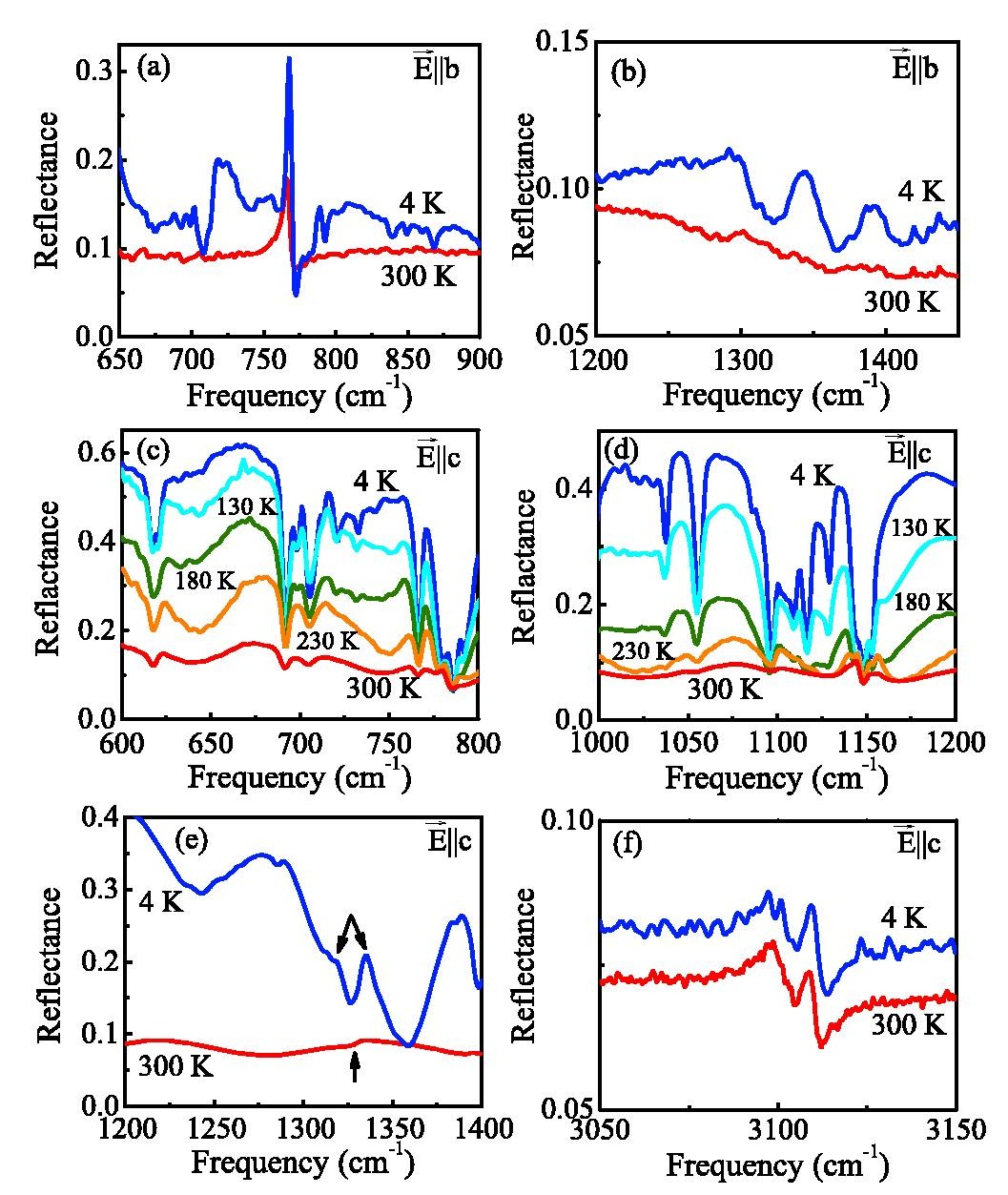}}
  \caption{\label{spectra}(Color online)
  Close-up views of several vibrational features of
(DTTTF)$_2$Cu(mnt)$_2$ in reflectance mode. (a) in-phase breathing
mode of the outer ring of DTTTF cation involving C-S stretch and
C=C-S bend, (b) symmetric C-C stretching mode of the DTTTF cation,
(c) C-S-C bending and C-S stretching modes of the DTTTF cation, (d)
asymmetric C-C + C-S stretch + C-H wag of DTTTF, S-C=C-H wagging
motion of the DTTTF cation, and C-C stretch of the Cu(mnt)$_2$
counterion, (e) out-of-phase C-C stretch in DTTTF, and (f) C-H stretching mode of the DTTTF cation. 
Measurement polarizations and temperatures are indicated in each
panel. The arrows in (e) denote features used in the charge
analysis.}
\end{figure}
C-S and  C-C stretching vibrations of the DTTTF cation based on our
dynamics calculations \cite{Wesolowski2003}. These modes are totally
symmetric and activated along the chain direction due to
electron-molecular vibrational coupling \cite{Dressel2004}. In the
$c$ direction, there are many infrared-active vibrations of $B_u$
symmetry \cite{Dressel2004}. Below, we focus on trends in the C-S-C
bending and C-S stretching features of the DTTTF cation between 600
and 800 cm$^{-1}$, the asymmetric C-C + C-S stretch + C-H wag and
the S-C=C-H wagging motion of the DTTTF cation
 between 1000 and
1200 cm$^{-1}$, the out-of-phase C-C stretch of the DTTTF cation at
$\sim$1335 cm$^{-1}$, and the C-H stretch of the DTTTF cation near
3100 cm$^{-1}$.

Figure \ref{spectra} displays close-up views of the reflectance  of
(DTTTF)$_2$Cu(mnt)$_2$  as a function of temperature.
As shown in Fig. \ref{spectra}(a) and (b), the  spectral response
above and
below 235 K  is consistent with a 
doubling along $b$ through the metal-insulator transition
\cite{Dias2004,Ribas2005}. There is also a striking
transverse distortion 
as evidenced by changes in the $c$-polarized reflectance spectra.
Analysis of the vibrational modes that probe the interchain
interactions  shows that symmetry breaking  involves the C-S bending
and stretching modes of the DTTTF cation,  combined C-C + C-S
stretching and C-H wagging, and S-C=C-H wagging perpendicular to the
chain (Fig. \ref{spectra}(c) and (d)). This result has important
implications for the intermolecular charge ordering pattern as
discussed below.

The variable temperature spectral response of (DTTTF)$_2$ Cu(mnt)$_2$
is different than that of the Au analog.
Two  separate
transitions are observed (220 and 70 K) in the latter.
The broad 220 K metal-insulator transition is driven by massive
symmetry breaking along the rung direction, whereas the 70 K
magnetic transition is associated with mode modifications in the
rail direction \cite{Wesolowski2003}. 
In contrast, the rung- and rail-directed lattice distortions are
coupled  in (DTTTF)$_2$Cu(mnt)$_2$. Both occur near 235 K,
consistent with the observation of a sharper transition. This may be
a consequence of stronger interchain interactions (see below).

We now discuss the implications of the combined $b$-axis doubling
and large transverse distortion at the metal-insulator transition
for the intermolecular charge ordering pattern in
(DTTTF)$_2$Cu(mnt)$_2$. Within the rectangular ladder model of Fig.
\ref{models}(d),
bond alternation along the rails is a necessity but charge ordering
is not anticipated  \cite{Yan2006}. The symmetry between the chains
implies that this model cannot account for the strong transverse
distortion in (DTTTF)$_2$Cu(mnt)$_2$: specifically, the symmetry
breaking involving C-S, C-C, and complex wagging motion
perpendicular to chain direction (Fig. \ref{spectra}(c) and (d)).
The proposed
Heisenberg spin ladder model with 
(DTTTF)$_2$$^{+.}$ dimers acting as the $S$ = $\frac{1}{2}$ sites (Fig. \ref{models}(c))
\cite{Rovira2001,Ribas2005} is therefore not applicable
 to this
class of materials.

The zigzag ladder model provides a more promising framework for
understanding intermolecular charge ordering in
(DTTTF)$_2$Cu(mnt)$_2$. The bond-charge density wave ground state
shown in Fig. \ref{models}(e) is a consequence of cooperative charge
and bond ordering \cite{Clay2005}. Predictions of the model include
a
large spin gap, 
charge ordering, and
\begin{figure}[tbh]
\centerline{
\includegraphics[width = 1.91 in]{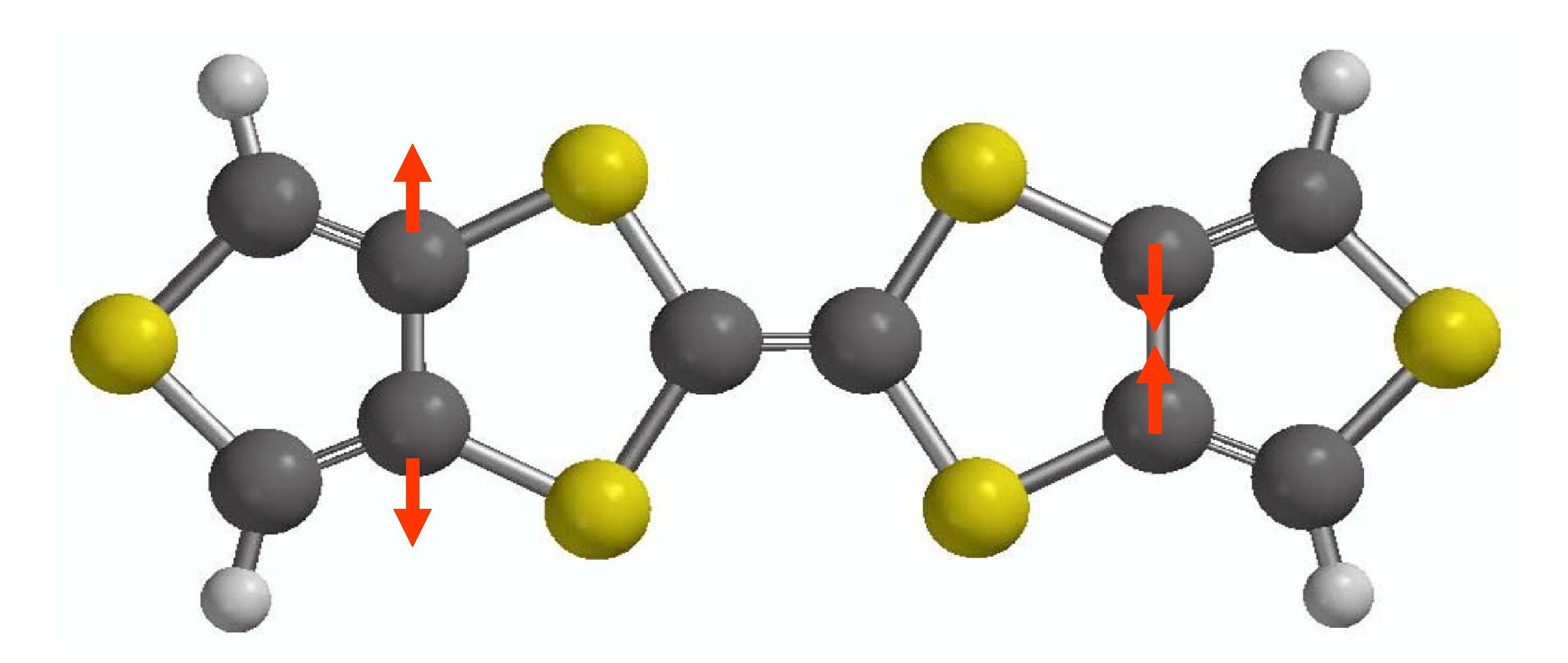}}
\caption{\label{mode} (Color online) Schematic  mode displacement
pattern for the out-of-phase C-C stretch of the DTTTF cation. }
\end{figure}
a complex structural modification involving both leg and diagonally
transverse distortions. The  strong $c$-polarized distortions
observed in the spectra (Fig. \ref{spectra}) are
consistent with these predictions. 
Indeed, the zigzag ladder model can potentially also explain the
occurrence of two distinct transitions in (DTTTF)$_2$Au(mnt)$_2$ but
single transition behavior in (DTTTF)$_2$Cu(mnt)$_2$. Recall that
the co-operative distortion of Fig.~1(e) requires a minimum
interchain hopping $t_{int} >$ 0.59$t$, where $t$ is the intrachain
one-electron hopping integral \cite{Clay2005}. We believe that
$t_{int}$ in (DTTTF)$_2$Au(mnt)$_2$ may be smaller than this
threshold value at high temperatures, and the nearly independent
stacks undergo the usual high temperature 4k$_F$ charge ordering
transition observed also in the quasi-one-dimensional TMTTF-based
$\frac{1}{4}$-filled band systems  \cite{Clay2007}.
The charge disproportionation is then exactly as in Fig.~1(e), but
all bonds are uniform. This would give rise to the observed
asymmetry between the stacks \cite{Wesolowski2003}. As the
temperature is lowered to below 70 K, either because of lattice
contraction or molecular rotation \cite{Mori},
$t_{int}$ is larger than the threshold value, the system behaves as
the zigzag ladder in the spin singlet subspace, and the co-operative
bond distortions of Fig.~1(e) occur. Within this scenario $t_{int}$
in the Cu-analog is already larger than 0.59$t$, which leads to the
single observed transition involving both charge and spin. Complete
theoretical proof of this scenario requires the demonstration of a
single transition in the $\frac{1}{4}$-filled band zigzag ladder
with large $t_{int}$ and work is currently in
progress. 

\begin{figure}[tbh]
\centerline{
\includegraphics[width = 3.3 in]{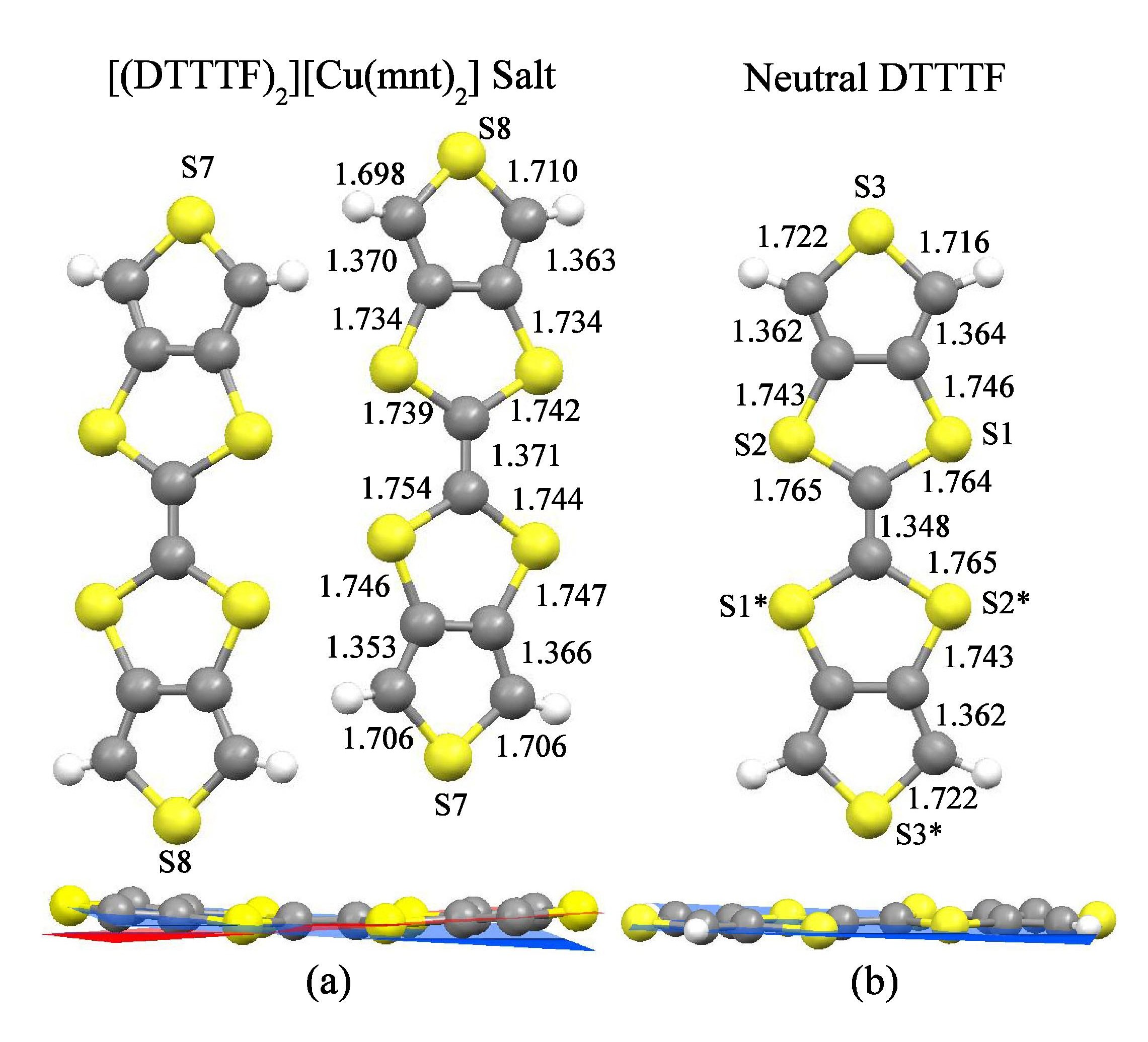}}
\caption{\label{structure} (Color online)(a) Bond lengths and
relative orientation of the two independent  cations in
(DTTTF)$_2$Cu(mnt)$_2$, and (b) those of neutral DTTTF in the
molecular
crystal at 300 K \cite{Rovira2001,Ribas2005}.  
The cations display an out of plane distortion in the salt (with a
5.78$^{\circ}$ angle between the thiophene rings and the TTF core)
in addition to strong molecular asymmetry. In the neutral compound,
DTTTF  is nearly flat, although still asymmetric; both molecules in
the unit cell are identical. Interestingly, the 300 K vibrational
spectrum of the neutral DTTTF molecular crystal (not shown) also
displays mode splitting that we take as an indication of
intramolecular charge inhomogeneity. Asymmetry in the local
molecular structure supports this assignment.
Distances are in \AA. 
}
\end{figure}

In order to evaluate the
molecular charges in the low temperature phase, we followed the
approach outlined by Yamamoto {\it et. al.}\cite{Yamamoto2005} and
selected an unperturbed, infrared-active vibrational mode that, from
our dynamics calculations \cite{Wesolowski2003}, is sensitive to
charge.  
Several authors have observed that vibrational frequencies depend
linearly on average molecular charge in the absence of strong
coupling \cite{Dressel2004}. The advantage of using an unperturbed
(rather than strongly coupled) mode for the analysis is that
complementary Raman data is not required \cite{Yamamoto2005}. Our
analysis focused on an infrared-allowed out-of-phase C-C stretch of
DTTTF (Fig. \ref{mode}). This mode is similar in character to
$\nu$$_{27}$ in the BEDT-TTF building block molecule
\cite{Yamamoto2005}. Accounting for the typical 3\% overestimate of
calculated frequencies and the flat vs. boat-shaped nature of the
neutral vs. cationic building block molecules (see Fig.
\ref{structure} and discussion, below), we compared the calculated
results and the associated linear fit ($\rho$ = -0.022$\omega$ +
29.82) with the experimental optical conductivity. From features at
$\sim$1319.5 and 1334.5 cm$^{-1}$ in $\sigma$$_1$($\omega$), we
estimate the difference in the site charges in the charge-ordered
low temperature phase to be $\sim$ 0.35 - 0.4.
These numbers 
should be compared with those obtained by Raman spectroscopy.
The absence of strong low temperature splitting of 
the C-H stretching mode on the cation (Fig. \ref{spectra}(f))
indicates that charge rich and charge poor regions are located at
the heart of the DTTTF building block molecule. 


Although not part of the aforementioned zigzag model for
intermolecular charge ordering, on-site charge asymmetry within the
DTTTF molecule is expected from the chemical and structural point of
view. Examination of the structure (Fig. \ref{structure}(a)) shows
that the DTTTF cations are locally distorted, with irregular bond
lengths (and angles) \cite{Rovira2001,Ribas2005}. This molecular
asymmetry combined with the boat-type distortion of the cation gives
rise to intramolecular charge inhomogeneities, where the local
concentration of charge is different from the average charge on an
individual molecule.
An inhomogeneous on-site
charge distribution is evident in the vibrational spectrum as
additional mode splitting (for instance, fine structure on the C-S
bending and stretching modes of the DTTTF cation as well as the
combined C-C + C-S stretch + C-H wag and the complex S-C=C-H wagging
motion of the DTTTF cation), overlaying the aforementioned
intermolecular charge ordering effects. Based on the small size of
the additional mode splitting, the local charge distribution
deviates only slightly  from the average 
distribution at room temperature. The fine structure is more
well-defined at low temperature, but the energy scale does not
change very much. We therefore estimate
$\mid$$\Delta$$\rho$$\mid$=$\mid$$\rho$$_{ave}$-$\rho$$_{loc}$$\mid$$\ll$0.1
with respect to the average 4 K charge distribution. Mode splitting
of a similar size ($\sim$10 cm$^{-1}$) has been observed in
(TTM-TTP)I$_3$ and attributed to an asymmetric deformation of the
organic building block molecule and consequent non-uniformity of the
intramolecular charge distribution \cite{Swietlik2005}.
Pressure may suppress the intermolecular charge ordering and on-site
charge inhomogeneities in (DTTTF)$_2$Cu(mnt)$_2$, leading to the
metallic state \cite{Dias2004}.

%

%
%

%

We measured the variable temperature infrared response of the spin
ladder candidate (DTTTF)$_2$Cu(mnt)$_2$ in order to assess the
intermolecular charge ordering pattern  within the context of two
recent recent theoretical predictions.
 The low temperature distortion along the chain direction
through the 235 K metal-insulator transition  is consistent with
expectations based on structural studies for a doubling along $b$.
At the same time, there is a large transverse
 distortion. The presence of this transverse
lattice distortion rules out rectangular ladder models and suggests
that the zigzag ladder model with its cooperative charge and bond
ordering may be more appropriate for
(DTTTF)$_2$Cu(mnt)$_2$. Comparison of these results with our
previous work on the Au analog material shows that the
metal-insulator transitions are quite different. Coupled chain- and
rung-directed distortions are observed in (DTTTF)$_2$Cu(mnt)$_2$,
whereas they are separate in the Au compound. We suspect that this
is a consequence of stronger interchain interaction in the Cu-based
system. A mode analysis provides microscopic information on the
vibrational processes driving the metal-insulator transition in
(DTTTF)$_2$Cu(mnt)$_2$. 

This research is supported by NSF-DMR-0600089 (UT), DOE
DE-FG02-06ER46315 (UA and MSU), DGI-CTQ2006-06333 (Spain),
POCI/QVI/57528/2004 (Portugal), and MAGMANET. We thank J.T.
Haraldsen, I. Olejniczak, and H. Mikre for useful discussions.

\vspace{0.08in} \noindent
 $^*$ Presently at University College
London.

\end{document}